\documentclass[showpacs,preprintnumbers,amsmath,amssymb,prb,superscriptaddress]{revtex4-1}
\usepackage{graphicx}

\usepackage{color}

\begin{document}

\title{Two-magnon excitations in resonant inelastic x-ray scattering studied by spin-density-wave formalism}

\author{Takuji~Nomura}
\email{nomurat@spring8.or.jp}
\affiliation{
Synchrotron Radiation Research Center, National Institutes for Quantum and Radiological Science and Technology, 
SPring-8, 1-1-1 Kouto, Sayo, Hyogo 679-5148, Japan}

\date{\today}

\begin{abstract}
We study two-magnon excitations in resonant inelastic x-ray scattering (RIXS) at the transition-metal $K$-edge. 
Instead of working with effective Heisenberg spin models, we work with a Hubbard-type model ($d$-$p$ model) 
for a typical insulating cuprate La$_2$CuO$_4$. 
For the antiferromagnetic ground state within the spin-density-wave (SDW) mean-field formalism, 
we calculate the dynamical correlation function within the random-phase approximation (RPA), 
and then obtain two-magnon excitation spectra by calculating the convolution of it. 
Coupling between the $K$-shell hole and the magnons in the intermediate state is calculated 
by means of diagrammatic perturbation expansion in the Coulomb interaction. 
Calculated momentum dependence of RIXS spectra agrees well with that of experiments. 
A notable difference from previous calculations based on the Heisenberg spin models is that RIXS spectra 
have a large two-magnon weight near the zone center, which may be confirmed by further careful high-resolution experiments. 
\end{abstract}

\maketitle

\noindent

\section{Introduction}

Resonant inelastic x-ray scattering (RIXS) at the transition-metal absorption edges is a promising powerful tool 
to detect various elementary excitations in strongly correlated electron systems~\cite{Ament2011,Ishii2013}. 
Particularly, the RIXS technique utilizing the transition-metal $K$- or $L$-edge x-rays 
can probe momentum dependence of electronic excitations such as charge~\cite{Hasan2000,Kim2002,Suga2005}, 
orbital~\cite{Ishii2011,Jarrige2012}, and magnon excitations~\cite{Braicovich2010,Guarise2010}. 
To analyze theoretically those momentum-dependent excitations in RIXS, 
various effective theoretical methods have been adopted, e.g., exact diagonalization~\cite{Tsutsui1999,Ide2000,Jia2016}, 
perturbation expansion\cite{Nomura2004,Nomura2005,Igarashi2006}, 
ultra-short life-time expansion~\cite{VanDenBrink2006,Ament2007}, 
dynamical mean-field theory (DMFT)~\cite{Pakhira2012}. 

In x-ray scattering with linearly polarized x-rays, the total spin moment is conserved, 
if effects of the spin-orbit coupling are negligible. 
Therefore, in contrast to neutron scattering, only an even number of magnons can be excited, 
whereas an odd number of magnons are prohibited to be excited. 
In fact, excitations with 500 meV energy and characteristic momentum dependence were observed in the Cu $K$-edge RIXS 
for La$_2$CuO$_4$ ~\cite{Hill2008,Ellis2010}, and have been identified as two-magnon excitations, 
based on the agreement with theoretical calculations~\cite{Nagao2007,VanDenBrink2007,Vernay2007,Donkov2007,Forte2008}. 
In those previous theoretical works they adopted Heisenberg spin Hamiltonians and the spin-wave (SW) approximation. 

A large number of theoretical studies on two-magnon excitations have been done in the context of Raman light scattering 
over several decades~\cite{Fleury1968a,Elliott1968,Fleury1968b, 
Elliott1969,Parkinson1969,Fleury1970,Singh1989,Shastry1990,Canali1992,Chubukov1995,Schonfeld1997,Sandvik1998}. 
The essential microscopic process in most of those studies is the inter-site spin exchange, 
which can be described effectively by the so-called Fleury-Loudon (FL) Hamiltonian~\cite{Fleury1968a}: 
\begin{equation}
H_{FL} = \alpha \sum_{ \langle i,j \rangle} 
[{\bf e} \cdot {\bf r}_{ij}][{\bf e}' \cdot {\bf r}_{ij}] {\bf s}_i \cdot {\bf s}_j, 
\label{eq:FL}
\end{equation}
where $\alpha$ is a constant, ${\bf s}_i$ is the spin operator at magnetic ion site $i$, 
${\bf e}$ and ${\bf e}'$ are the electric-field vectors of emitted and absorbed rays, 
and ${\bf r}_{ij}$ is the coordinate vector connecting sites $i$ and $j$. 
The FL Hamiltonian can be incorporated into Heisenberg spin Hamiltonians as perturbation with ease. 
In calculations with the FL Hamiltonian, one can reproduce the observed characteristic lineshapes 
of two-magnon Raman spectra by taking account of magnon-magnon interactions, which crucially evidenced 
the importance of magnon-magnon interactions~\cite{Elliott1968,Fleury1968b,Elliott1969,Parkinson1969,Fleury1970}. 
The FL Hamiltonian has been derived microscopically from a Hubbard Hamiltonian in the large-$U$ limit, 
being intended for the high-$T_c$ cuprates~\cite{Shastry1990,Chubukov1995}. 

Two-magnon excitations in RIXS should be distinguished from those in Raman light scattering, 
since a $K$-shell (i.e., $1s$ core) hole created in the intermediate state can play an essential role. 
We need to consider excitation processes in the presence of the core hole at a transition-metal site, 
which are not involved in Raman light scattering. 
A microscopic mechanism of two-magnon excitations in $K$-edge RIXS was before proposed by van den Brink~\cite{VanDenBrink2007}. 
He took account of virtual inter-site hopping processes to screen the $1s$ hole in the intermediate state of RIXS, 
and thereby calculated the modification to the antiferromagnetic Heisenberg spin exchange $J$ around the $1s$ hole. 
The modified exchange integral between the excited site and neighboring sites induces 
inter-site spin exchange excitations between those sites, and consequently two magnons are 
created before the $1s$ hole is finally annihilated. 
In Ref.~\onlinecite{Nagao2007}, Nagao and Igarashi incorporated this mechanism into the previous perturbative framework 
developed by the author and Igarashi. 
They replaced a ladder of electron and hole propagators in Ref.~\onlinecite{Nomura2005} with a ladder of magnon propagators, 
and calculated the dynamical correlation function by using the SW approximation 
and including the magnon-magnon interaction within $1/S$ expansion. 
These calculations seem to explain well the  experimental observations so far 
in La$_2$CuO$_4$~\cite{Nagao2007,VanDenBrink2007,Forte2008}. 

A central aim of the present work is to describe two-magnon excitations in $K$-edge RIXS 
using an itinerant Hubbard-type model ($d$-$p$ model) and the spin-density-wave (SDW) formalism, 
instead of using a Heisenberg spin model and the SW approximation. 
To our knowledge, studies of two-magnon RIXS based on the SDW formalism are still lacking, 
although some studies of two-magnon Raman light scattering based on the SDW formalism have been done~\cite{Schonfeld1997}. 
We deal with the coupling between the $1s$ hole and a pair of magnons within perturbation expansion in the Coulomb interaction, 
and calculate the magnon propagators within the random-phase approximation (RPA). 
This approach is another natural way of extending our previous perturbative formulation 
to explain well the two-magnon RIXS spectra, as we shall demonstrate below. 

\section{Theoretical Framework}

\subsection{Schematic illustration of two-magnon RIXS and Hamiltonian}

\begin{figure}[htb]
\includegraphics[width=0.7\columnwidth]{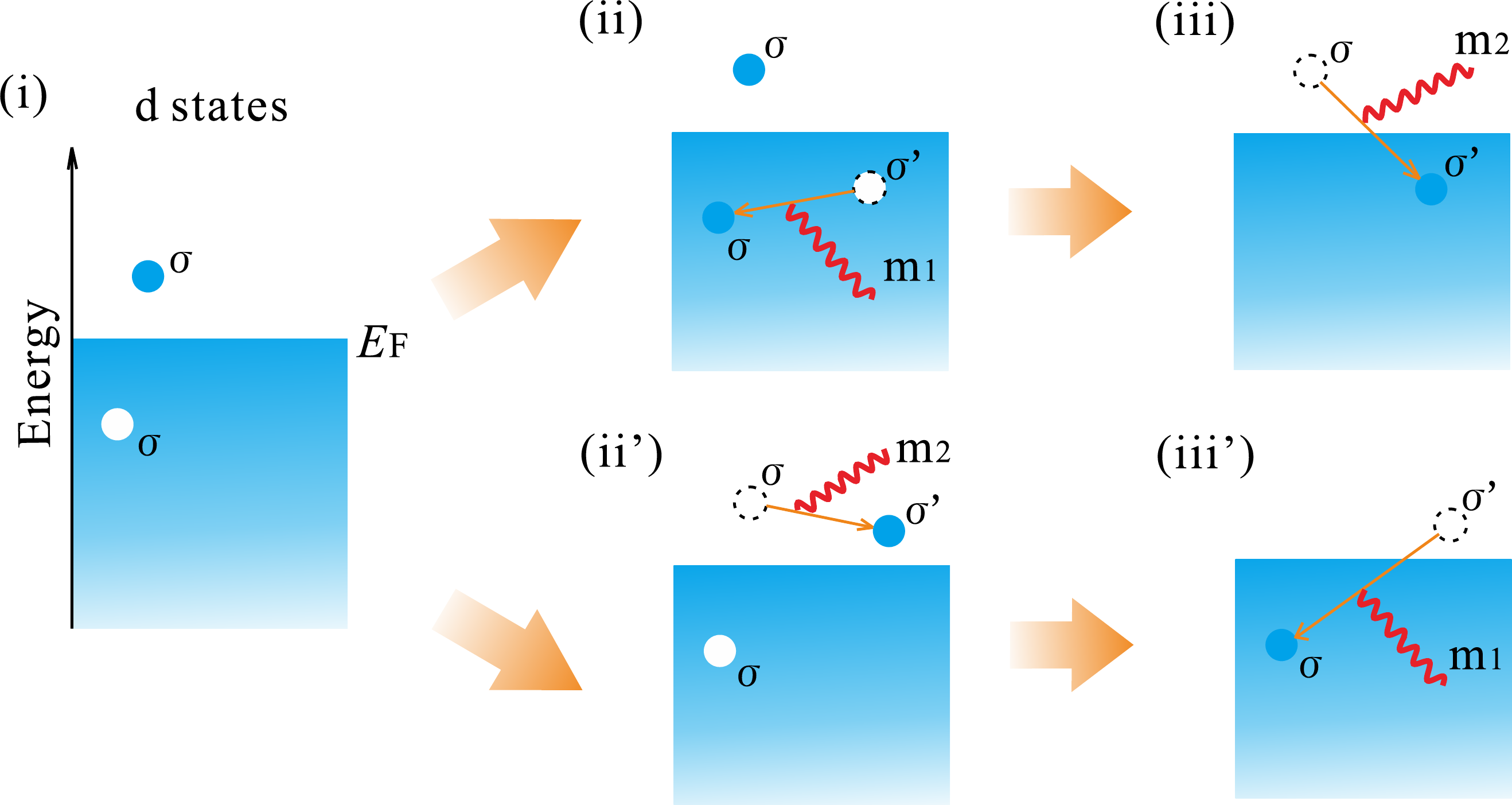}
\caption{
(Color online)
Schematic picture of typical processes of two-magnon creation. 
(i) An electron-hole pair is excited on the $d$ bands to screen the $1s$ core hole 
(Only the correlated $d$-electron states are drawn, and the excited $1s$ hole and $p$ electron 
are not shown explicitly). $\sigma$ represents a spin state. 
(ii) An electron below $E_F$ goes into the created hole state, 
by changing its spin ($\sigma' \rightarrow \sigma$) and exciting a magnon m$_1$. 
Wavy line represents a magnon. 
(iii) Finally, the electron excited initially above $E_F$ goes into the hole state below $E_F$ 
by changing its spin ($\sigma \rightarrow \sigma'$) and exciting another magnon m$_2$. 
Another typical process of two-magnon creation is represented 
by (i) $\rightarrow$ (ii') $\rightarrow$ (iii'). }
\label{fig1}
\end{figure}

Before presenting theoretical details, we schematically illustrate typical $K$-edge RIXS processes involving two-magnon creation: 
An incident x-ray near the $K$-edge promotes a $1s$ electron resonantly to an empty $p$ band above the Fermi energy ($E_F$). 
Subsequently, correlated $d$ electrons near $E_F$ are excited to screen the created $1s$ hole, 
due to the Coulomb interaction between the $1s$ and $d$ orbitals. 
Such a screening process can be expressed by excitation of an electron-hole pair, as illustrated in Fig.~\ref{fig1} (i). 
This electron-hole excitation in the intermediate state can decay into a pair of magnons, 
due to the spin degree of freedom of the excited electron and hole, as shown in Fig.~\ref{fig1} (ii) and (iii). 
After two magnons are created, the initially-promoted $p$ electron goes back to the $1s$ state, emitting an x-ray, in the final state. 
The energy loss and momentum change between the incident and emitted x-rays are carried away by the two magnons. 

To be specific, hereafter we restrict our discussion to a typical copper oxide La$_2$CuO$_4$, 
although our discussion below is applicable also to other transition-metal compounds. 
To describe the electron dynamics illustrated above, we use the following Hamiltonian: 
\begin{equation}
H = H_d + H_s + H_{sd} + H_p  + H_x. 
\end{equation}
This is the same Hamiltonian that we used in Ref.~\onlinecite{Nomura2005}.
We present each term explicitly in the following. 

$H_d$ describes the electronic states near $E_F \equiv 0$. 
We take a Hubbard-type Hamiltonian ($d$-$p$ model) for the Cu$3d_{x^2-y^2}$ and O$2p_{x,y}$ orbitals in a single CuO$_2$ layer: 
\begin{eqnarray}
H_d &=& H_0 + H', \\
H_0 &=& \sum_i \sum_\sigma \varepsilon_d d_{i \sigma}^{\dag} d_{i \sigma} 
+ \sum_a \sum_{\ell=x,y} \sum_\sigma \varepsilon_p p_{a \ell \sigma}^{\dag} p_{a \ell \sigma} \nonumber\\
&&+ \sum_{\langle i,a \rangle} \sum_{\ell=x,y}  \sum_\sigma t_{dp} (p_{a \ell \sigma}^{\dag} 
d_{i \sigma} + d_{i \sigma}^{\dag} p_{a \ell \sigma} ) 
+ \sum_{\langle a,b \rangle} \sum_\sigma t_{pp} (p_{a x \sigma}^{\dag} p_{b y \sigma} 
+ p_{b y \sigma}^{\dag} p_{a x \sigma}), \\
H' &=& \frac{1}{2} \sum_i \sum_{\sigma \neq \sigma'} U n_{d i \sigma} n_{d i \sigma'}. 
\end{eqnarray}
Here $d_{i\sigma}$ and $p_{a\ell\sigma}$ ($d_{i\sigma}^{\dag}$ and $p_{a\ell\sigma}^{\dag}$) 
are the annihilation (creation) operators for the Cu$3d_{x^2-y^2}$ and O$2p_{\ell}$ electrons with spin $\sigma$, 
where the Cu$3d_{x^2-y^2}$ and O$2p_\ell$ orbitals form a $\sigma$-bond. 
$n_{di\sigma}$ is the number operator for the $3d_{x^2-y^2}$ electrons with spin $\sigma$ at Cu site $i$. 
Summation with $\langle i,a \rangle$ ($\langle a,b \rangle$) is over nearest-neighbor Cu-O (O-O) bonds. 
We take $t_{dp} = 1.3$ eV, $t_{pp} = 0.65$ eV~\cite{Hybertsen1989}, and $U=11$ eV 
as in our previous work~\cite{Nomura2005}. 

For the $1s$ electrons, we assume a completely localized orbital at each Cu site: 
\begin{equation}
H_s = \sum_i \sum_{\sigma} \varepsilon_{1s} s_{i\sigma}^{\dag} s_{i\sigma} 
= \sum_{\bf k} \sum_{\sigma} \varepsilon_{1s} s_{{\bf k}\sigma}^{\dag} s_{{\bf k}\sigma}, 
\end{equation}
where $\varepsilon_{1s}$ is the one-particle energy, $s_{i\sigma}$ ($s_{i\sigma}^{\dag}$) 
is the annihilation (creation) operator for the $1s$ electrons with spin $\sigma$ at Cu site $i$. 
$ s_{{\bf k}\sigma} (s_{{\bf k}\sigma}^{\dag})$ is the momentum representation 
of $ s_{i \sigma} (s_{i \sigma}^{\dag})$. 

$H_{sd}$ is the core-hole potential, i.e., the Coulomb interaction between the $1s$ and $3d$ electrons: 
\begin{eqnarray}
H_{sd} &=& V_{sd} \sum_i \sum_{\sigma \sigma'} 
s_{i\sigma}^{\dag} s_{i\sigma} d_{i\sigma'}^{\dag} d_{i\sigma'} \nonumber\\
&=& V_{sd} \sum_i n_{si} n_{di}, 
\label{eq:Hsd}
\end{eqnarray}
where $n_{di}$ and $n_{si}$ are the number operator for the $3d_{x^2-y^2}$ and $1s$ electrons 
with spin $\sigma$ at Cu site $i$, respectively. 
Since the $3d$ and inner-shell $1s$ orbitals are both strongly localized, 
the core-hole potential $V_{sd}$ is usually chosen to be comparable to the Cu$3d$ Coulomb interaction $U$ 
for La$_2$CuO$_4$. 

$H_p$ describes the conduction $4p$ band electrons: 
\begin{equation}
H_p = \sum_{\bf k} \sum_{\mu} \sum_{\sigma} \varepsilon_{4p \mu}({\bf k}) p'{}_{{\bf k}\mu\sigma}^{\dag} p'{}_{{\bf k} \mu \sigma}, 
\end{equation}
where $p'{}_{{\bf k} \mu \sigma}$ ($p'{}_{{\bf k} \mu \sigma}^{\dag}$) is the annihilation (creation) operator 
of the $4p_{\mu}$ electron ($\mu=x,y,z$) with momentum ${\bf k}$ and spin $\sigma$. 
$H_x$ describes the resonant $1s$-$4p$ electric-dipole transition induced by x-rays: 
\begin{equation}
H_x = \sum_{\bf{k},\bf{q}} \sum_{\mu} \sum_{\sigma} 
w_{\mu}({\bf q},{\bf e}) \alpha_{\bf{q}\bf{e}} p'{}_{{\bf k}+{\bf q} \mu \sigma}^{\dag} s_{{\bf k}\sigma} + h.c., 
\end{equation} 
where $\alpha_{\bf{q}\bf{e}}$ is the annihilation operator of a photon 
with momentum $\bf{q}$ and polarization $\bf{e}$. 
The electric-dipole transition matrix $w_{\mu}(\bf{q}, \bf{e})$ is given by 
\begin{equation}
w_{\mu}({\bf q} , {\bf e}) = - \frac{e}{m}\sqrt{\frac{2\pi}{|{\bf q}|}} {\bf e} 
\cdot \langle 4p_{\mu} |{\bf p}|1s \rangle \propto {\bf e} \cdot {\bf e}_\mu, 
\end{equation}
in natural units ($c = \hbar = 1$), where $e$ and $m$ are the elementary charge and the electron mass, 
${\bf e}_\mu$'s are the orthonormal basis vectors. 

\subsection{Two-magnon RIXS formula}
\label{sbsc:Two-magnon RIXS formula}

Nozi\`{e}res and Abrahams (NA) developed a theoretical framework of electron Raman scattering 
by means of Keldysh perturbation theory, and discussed the threshold singularity in metals~\cite{Nozieres1974}. 
Extending the NA's framework, we derived a formula for RIXS intensity 
at the transition-metal $K$ edge~\cite{Nomura2004,Nomura2005,Igarashi2006}, and thereby we explained experimental observations 
on charge-transfer and orbital excitations in $K$-edge RIXS for several transition-metal compounds~\cite{Takahashi2007,Semba2008,Nomura2014}. 
To calculate two-magnon RIXS spectra, we further extend our previous framework by a way different from 
Ref.~\onlinecite{Nagao2007}, fully based on the itinerant picture. 

We adopt the SDW mean-field approach to describe the antiferromagnetic (AF) ground state~\cite{Schrieffer1989,Peres2002}. 
Within the SDW mean-field formalism, the Coulomb interaction part $H'$ is approximated by 
\begin{eqnarray}
H'_{MF} &=& \frac{U}{2} \sum_{\bf k} \sum_{\sigma_i} 
d_{{\bf k}_{\sigma_1} \sigma_1}^{\dag} [
n_d \delta_{\sigma_1\sigma_2} - {\bf m} \cdot {\bf \sigma}_{\sigma_1\sigma_2} ]
d_{{\bf k}_{\sigma_2} \sigma_2} \nonumber\\ 
&& - \frac{NU}{4} [ n_d^2 - |{\bf m}|^2 ], 
\end{eqnarray}
where ${\bf k}_\sigma$ is defined as ${\bf k}_\uparrow = {\bf k}$ and ${\bf k}_\downarrow = {\bf k} + {\bf q}_{AF}$ 
with ${\bf q}_{AF}$ the magnetic-ordering vector. 
$n_d$ and ${\bf m}$ are the mean-fields for the $d$-electron number 
and spin moment to be determined self-consistently, 
and ${\bf \sigma}$ is the Pauli matrix vector. 
In our calculation, we set the $z$-axis of spin along the crystallographic [001] (i.e., the $c$-axis) direction, 
and assume the commensurate AF ground state with ${\bf q}_{AF}=(\pi,\pi)$ 
and ${\bf m} \parallel [110]$ as observed in neutron scattering~\cite{Vaknin1987}. 
Introducing new fermion annihilation and creation operators, 
the mean-field Hamiltonian $H_{d,MF} \equiv H_0 + H'_{MF}$ is diagonalized as: 
\begin{equation}
H_{d,MF} = \sum_{\bf k} \sum_j E_j({\bf k}) a_{j{\bf k}}^{\dag} a_{j{\bf k}}, 
\end{equation}
where $j$ is band index. $E_j({\bf k})$ is the diagonalized band energy which the chemical potential is already subtracted from. 
The chemical potential is always determined so that the total electron number equals five per unit cell. 
The original $d$-electron annihilation and creation operators in the momentum representation 
are related to $a_{j{\bf k}}$ and $a_{j{\bf k}}^{\dag}$ by 
\begin{eqnarray}
d_{{\bf k}_{\sigma} \sigma} &=& \sum_j u_{\sigma,j}({\bf k}) a_{j{\bf k}}, \\
d_{{\bf k}_{\sigma} \sigma}^{\dag} &=& \sum_j u_{\sigma,j}^*({\bf k}) a_{j{\bf k}}^{\dag}, 
\end{eqnarray}
where $u_{\sigma,j}({\bf k}_{\sigma})$'s are the $d$-electron elements of the diagonalization matrix. 
The $d$-electron Green's function has a $2 \times 2$ matrix form: 
\begin{equation}
\hat{G}({\bf k}, \xi) = \left[
\begin{array}{cc}
G_{\uparrow\uparrow}({\bf k}, \xi)   & G_{\uparrow\downarrow}({\bf k}, \xi) \\
G_{\downarrow\uparrow}({\bf k}, \xi) & G_{\downarrow\downarrow}({\bf k}, \xi)
\end{array}
\right]. 
\end{equation}
Each element of $\hat{G}({\bf k}, \xi)$ is expressed as 
\begin{equation}
G_{\sigma_1\sigma_2}({\bf k}, \xi) 
= \sum_j u_{\sigma_1,j}({\bf k}) u_{\sigma_2,j}^*({\bf k}) G_j({\bf k},\xi), 
\end{equation}
with 
\begin{equation}
G_j({\bf k},\xi) = \frac{1}{\xi - E_j({\bf k})}. 
\end{equation}
The advanced, retarded and causal branches of the Green's functions are 
\begin{eqnarray}
G^R_j({\bf k}, \xi) &\equiv & G_j({\bf k}, \xi + i \gamma ) = \frac{1}{\xi - E_j({\bf k}) + i \gamma}, \\
G^A_j({\bf k}, \xi) &\equiv & G_j({\bf k}, \xi - i \gamma ) = \frac{1}{\xi - E_j({\bf k}) - i \gamma}, \\
G^C_j({\bf k}, \xi) &\equiv & [1-f(\xi)] G^R_j({\bf k}, \xi) + f(\xi) G^A_j({\bf k}, \xi), 
\end{eqnarray}
where $\gamma$ is a small positive, and $f(\xi) = [e^{\xi/T} + 1]^{-1} $ 
is the Fermi distribution function at temperature $T$. 
Using the Green's function $G^R_j({\bf k}, \xi)$,
the self-consistency equations for $n_d$ and ${\bf m}$ become 
\begin{eqnarray}
n_d &=& - \sum_j \frac{1}{N} \sum_{\bf k} \sum_{\sigma_i} 
\int_{-\infty}^{\infty} \frac{d \xi}{\pi} 
u_{\sigma_1,j}^*({\bf k}) \delta_{\sigma_1\sigma_2} u_{\sigma_2,j}({\bf k}) \nonumber \\
&& \times f(\xi) {\rm Im}[G^R_j({\bf k},\xi)], \\
{\bf m} &=& - \sum_j \frac{1}{N} \sum_{\bf k} \sum_{\sigma_i} 
\int_{-\infty}^{\infty} \frac{d \xi}{\pi} 
u_{\sigma_1,j}^*({\bf k}) {\bf \sigma}_{\sigma_1\sigma_2} u_{\sigma_2,j}({\bf k}) \nonumber \\
&& \times f(\xi) {\rm Im}[G^R_j({\bf k},\xi)]. 
\end{eqnarray}
Throughout our study, numerical integration in energy and momentum 
is carried out by discretizing the interval $-40$ eV $< \xi < 40$ eV into 16000 energy points 
and the first Brillouin zone into $N = 80 \times 80$ {\bf k} points. We take $\gamma = 0.16$ eV. 
We choose $\varepsilon_d - \varepsilon_p \approx -7 {\rm eV}$ 
so that $n_d = 1.53$ and $\varepsilon_d - \varepsilon_p + U n_d/2 = 1.4$ eV for the non-magnetic state. 
The spin moment and insulating gap calculated for the AF ground state 
are $|{\bf m}| = 0.61 \mu_B$ and 2.55 eV, respectively. 
The charge-transfer energy between the Cu-$3d$ and O-$2p$ states is 
\begin{equation}
\Delta = \varepsilon_d + U - \varepsilon_p \approx 4 \,{\rm eV}, 
\end{equation}
and thus the AF Heisenberg exchange between nearest-neighbor Cu sites is evaluated as~\cite{Zhang1988}: 
\begin{equation}
J \approx \frac{4t_{dp}^4}{\Delta^2} \biggl[ \frac{1}{U} + \frac{1}{2 \Delta} \biggr] \approx 0.15 \,{\rm eV}. 
\end{equation}
This value is consistent with that used in most Heisenberg-model calculations for La$_2$CuO$_4$. 

\begin{figure}[htb]
\includegraphics[width=0.4\columnwidth]{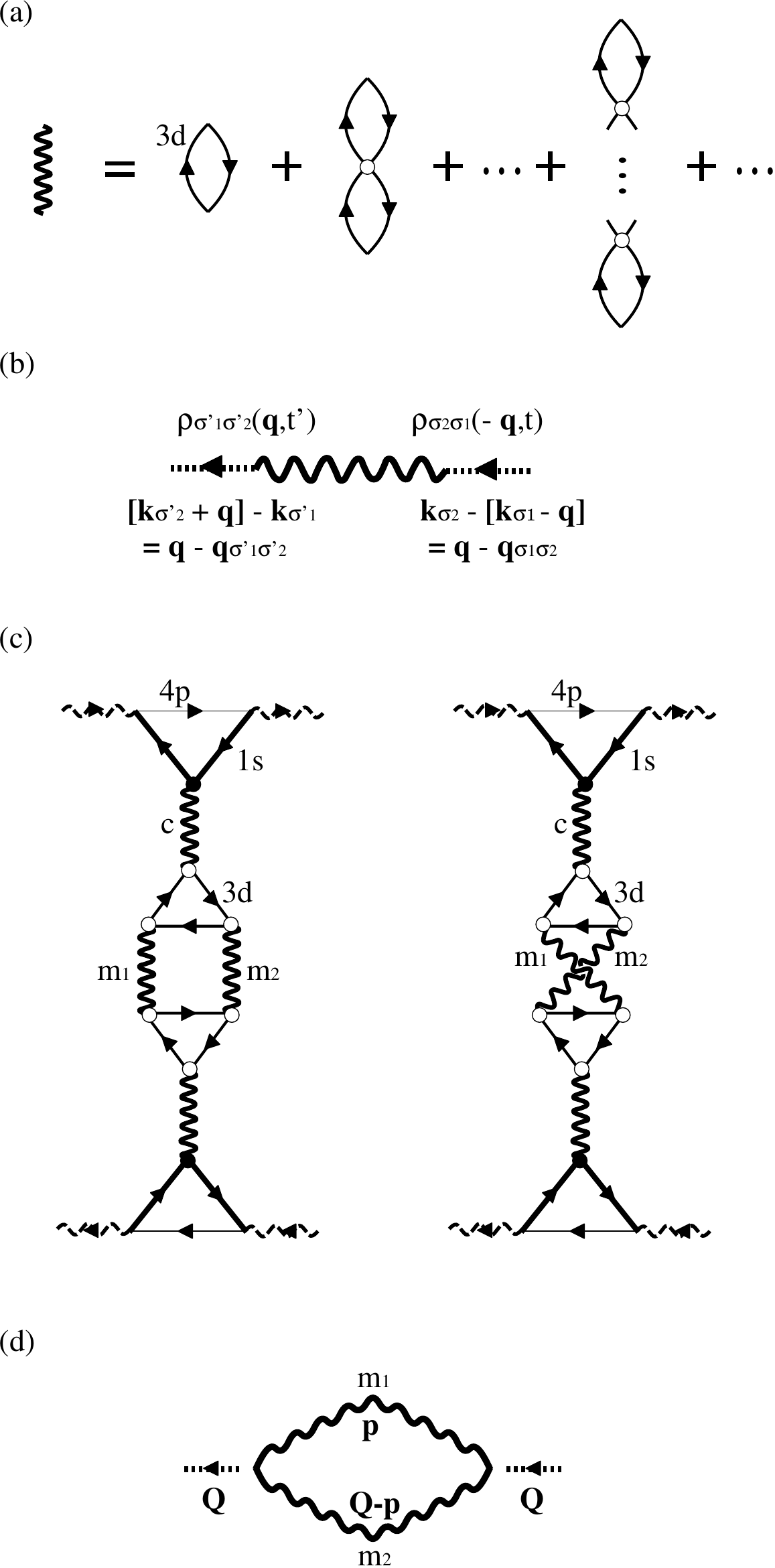}
\caption{
(a) Summation of diagrams for RPA. Thick wavy line and oriented solid lines represent the RPA susceptibility 
(i.e., the single-magnon propagator) and the Green's function for the 3$d$ electrons, respectively. 
The empty circle represents the 3$d$ Coulomb interaction matrix $\Gamma^{(0)}$. 
(b) Diagrammatic expression of $\Pi_{\sigma'_1\sigma'_2,\sigma_2\sigma_1}({\bf q}, t'-t)$, 
where the momenta absorbed to and emitted from the electron system 
are ${\bf q} - {\bf q}_{\sigma_1\sigma_2}$ and ${\bf q} - {\bf q}_{\sigma'_1\sigma'_2}$, respectively. 
(c) Typical two contributions from two-magnon excitations to RIXS spectra. 
External oriented broken wavy lines represent x-rays. 
Oriented thick, intermediate and thin solid lines represent the propagation 
of Cu-1$s$, 3$d$ and 4$p$ electrons, respectively. 
Filled circular vertex represents the Coulomb interaction $V_{sd}$ between the Cu-1$s$ and 3$d$ electrons at each Cu site. 
The thick solid wavy lines, c, m$_1$ and m$_2$, represent the RPA propagators, 
$\chi_{\tau_1\tau_2,\sigma\sigma}^C({\bf Q}, \Omega)$, 
$\Pi_{\sigma'_1\sigma'_2,\sigma_2\sigma_1}({\bf p}, \zeta)$ and 
$\Pi_{\sigma'_4\sigma'_3,\sigma_3\sigma_4}({\bf Q}-{\bf p}, \Omega-\zeta)$, respectively. 
(d) Extracted bare part of two-magnon propagation. 
The total carried momentum is ${\bf Q}$.}
\label{fig2}
\end{figure}

To describe the electron dynamics in RIXS, we define the dynamical correlation function by 
\begin{equation}
\Pi_{\sigma'_1\sigma'_2,\sigma_2\sigma_1}({\bf q}, t'-t) 
= \langle \rho_{\sigma'_1\sigma'_2}({\bf q},t') \rho_{\sigma_2\sigma_1}(-{\bf q},t) \rangle, 
\end{equation}
where $\rho_{\sigma_1\sigma_2}({\bf q},t)$ is the density operator in the Heisenberg representation:  
\begin{eqnarray}
\rho_{\sigma_1\sigma_2}({\bf q},t) &=& e^{i H_d t} \rho_{\sigma_1\sigma_2}({\bf q}) e^{-i H_d t}, \\
\rho_{\sigma_1\sigma_2}({\bf q}) &=& 
\sum_{\bf k} d_{{\bf k}_{\sigma_1}\sigma_1}^{\dag} d_{{\bf k}_{\sigma_2}+{\bf q}\sigma_2}. 
\end{eqnarray}
Fourier transform of the dynamical correlation function is related to the linear-response susceptibility 
in terms of the fluctuation-dissipation theorem: 
\begin{equation}
\Pi_{\sigma'_1\sigma'_2,\sigma_2\sigma_1}({\bf q}, \omega) = 
\frac{\chi_{\sigma'_1\sigma'_2,\sigma_2\sigma_1}^R({\bf q}, \omega) 
- \chi_{\sigma'_1\sigma'_2,\sigma_2\sigma_1}^A({\bf q}, \omega)}{i(1-e^{-\omega/T})}, 
\end{equation}
where $\chi_{\sigma'_1\sigma'_2,\sigma_2\sigma_1}^{R,A}({\bf q}, \omega)$ 
are the retarded and advanced branches of the susceptibility, and are the Fourier transforms of 
\begin{eqnarray}
\chi_{\sigma'_1\sigma'_2,\sigma_2\sigma_1}^R({\bf q}, t'-t) 
&=& i \theta(t'-t) \langle [ \rho_{\sigma'_1\sigma'_2}({\bf q},t'), \rho_{\sigma_2\sigma_1}(-{\bf q},t)] \rangle, \\
\chi_{\sigma'_1\sigma'_2,\sigma_2\sigma_1}^A({\bf q}, t'-t) 
&=& - i \theta(t-t') \langle [ \rho_{\sigma'_1\sigma'_2}({\bf q},t'), \rho_{\sigma_2\sigma_1}(-{\bf q},t)] \rangle. 
\end{eqnarray}
We calculate $\chi_{\sigma_1\sigma_2,\sigma_3\sigma_4}^{R,A}({\bf q}, \omega)$ within RPA: 
\begin{eqnarray}
\chi_{\sigma_1\sigma_2,\sigma_3\sigma_4}^{R,A}({\bf q}, \omega) = 
\chi_{\sigma_1\sigma_2,\sigma_3\sigma_4}^{(0)\,R,A}({\bf q}, \omega) \hspace{25mm}\nonumber\\
- \sum_{\sigma'_i} 
\chi_{\sigma_1\sigma_2,\sigma'_3\sigma'_4}^{(0)\,R,A}({\bf q}, \omega) 
\Gamma_{\sigma'_3\sigma'_1,\sigma'_2\sigma'_4}^{(0)} 
\chi_{\sigma'_1\sigma'_2,\sigma_3\sigma_4}^{R,A}({\bf q}, \omega), 
\end{eqnarray}
where 
\begin{equation}
\Gamma_{\sigma'_3\sigma'_1,\sigma'_2\sigma'_4}^{(0)} = 
U(\delta_{\sigma'_3\sigma'_4} \delta_{\sigma'_1\sigma'_2} - \delta_{\sigma'_3\sigma'_2} \delta_{\sigma'_1\sigma'_4}) 
\label{eq:coulomb}
\end{equation}
is the Coulomb interaction at Cu site. Diagrammatic expression of RPA is presented in Fig.~\ref{fig2}(a). 
$\chi_{\sigma_1\sigma_2,\sigma_3\sigma_4}^{(0)}({\bf q}, \omega)$ is the bare susceptibility calculated by 
\begin{eqnarray}
\chi_{\sigma_1\sigma_2,\sigma_3\sigma_4}^{(0)\,R}({\bf q}, \omega) = \frac{1}{N} \sum_{\bf k} 
\int_{-\infty}^{\infty} \frac{d \xi}{\pi} \sum_{j,j'} u_{\sigma_4,j}({\bf k}) u_{\sigma_1,j}^*({\bf k}) 
u_{\sigma_2,j'}({\bf k}+{\bf q}) u_{\sigma_3,j'}^*({\bf k}+{\bf q}) \nonumber\\
\times \Bigl[ f(\xi) {\rm Im} \{ G_j^R({\bf k},\xi) \} G_{j'}^R({\bf k}+{\bf q},\xi+\omega) 
+ f(\xi+\omega) G_j^A({\bf k},\xi) {\rm Im} \{ G_{j'}^R({\bf k}+{\bf q},\xi+\omega) \} \Bigr], \\
\chi_{\sigma_1\sigma_2,\sigma_3\sigma_4}^{(0)\,A}({\bf q}, \omega) = - \frac{1}{N} \sum_{\bf k} 
\int_{-\infty}^{\infty} \frac{d \xi}{\pi} \sum_{j,j'} 
u_{\sigma_4,j}({\bf k}) u_{\sigma_1,j}^*({\bf k}) 
u_{\sigma_2,j'}({\bf k}+{\bf q}) u_{\sigma_3,j'}^*({\bf k}+{\bf q}) \nonumber\\
\times \Bigl[ f(\xi) {\rm Im} \{ G_j^A({\bf k},\xi) \} G_{j'}^A({\bf k}+{\bf q},\xi+\omega) 
+ f(\xi+\omega) G_j^R({\bf k},\xi) {\rm Im} \{ G_{j'}^A({\bf k}+{\bf q},\xi+\omega) \} \Bigr]. 
\end{eqnarray}
For numerical calculations, we use more convenient expressions: 
\begin{eqnarray}
\chi_{\sigma_1\sigma_2,\sigma_3\sigma_4}^{(0)\,R}({\bf q}, \omega) &=& 
X_{\sigma_1\sigma_2,\sigma_3\sigma_4}^{(0)}({\bf q}, \omega) 
+ [X_{\sigma_2\sigma_1,\sigma_4\sigma_3}^{(0)}(- {\bf q}, - \omega)]^*, \\
\chi_{\sigma_1\sigma_2,\sigma_3\sigma_4}^{(0)\,A}({\bf q}, \omega) &=& 
[X_{\sigma_4\sigma_3,\sigma_2\sigma_1}^{(0)}({\bf q}, \omega)]^*
+ X_{\sigma_3\sigma_4,\sigma_1\sigma_2}^{(0)}(-{\bf q}, -\omega), 
\end{eqnarray}
where 
\begin{eqnarray}
X_{\sigma_1\sigma_2,\sigma_3\sigma_4}^{(0)}({\bf q}, \omega) &=& \frac{1}{N} \sum_{\bf k} 
\int_{-\infty}^{\infty} \frac{d \xi}{\pi} 
\Bigl[ f(\xi) \sum_j u_{\sigma_4,j}({\bf k}) u_{\sigma_1,j}^*({\bf k}) {\rm Im} \{ G_j^R({\bf k},\xi) \} \Bigr] \nonumber\\
&& \times \Bigl[ \sum_{j'} u_{\sigma_2,j'}({\bf k}+{\bf q}) u_{\sigma_3,j'}^*({\bf k}+{\bf q})
G_{j'}^R({\bf k}+{\bf q},\xi+\omega) \Bigr] . \label{eq:chi0b}
\end{eqnarray}
Energy-momentum integration in Eq.~(\ref{eq:chi0b}) is numerically performed using fast Fourier transformation (FFT). 

Below we need also the causal susceptibility, $\chi_{\sigma'_1\sigma'_2,\sigma_2\sigma_1}^C({\bf q}, \omega)$, 
which is the Fourier transform of 
\begin{equation}
\chi_{\sigma'_1\sigma'_2,\sigma_2\sigma_1}^C({\bf q}, t'-t) 
= i \langle {\rm T} [\rho_{\sigma'_1\sigma'_2}({\bf q},t') \rho_{\sigma_2\sigma_1}(-{\bf q},t)] \rangle. 
\end{equation}
Here T$[...]$ means the ordinary time-ordered product. 
This causal component can be calculated from the advanced and retarded components: 
\begin{equation}
\chi_{\sigma_1\sigma_2,\sigma_3\sigma_4}^C({\bf q}, \omega) = 
[b(\omega) + 1] \chi_{\sigma_1\sigma_2,\sigma_3\sigma_4}^R({\bf q}, \omega) 
- b(\omega) \chi_{\sigma_1\sigma_2,\sigma_3\sigma_4}^A({\bf q}, \omega), 
\end{equation}
where $b(\omega) = [e^{\omega/T}-1]^{-1} $ is the Bose distribution function. 

Single-magnon spectrum to be observed in neutron scattering is not 
$\Pi_{\sigma'_1\sigma'_2,\sigma_2\sigma_1}({\bf q}, \omega)$ itself, but 
\begin{equation}
S_{\mu\nu}({\bf q}, \omega) = 
\sum_{\sigma_i,\sigma'_i}[\sigma_\mu]_{\sigma'_1\sigma'_2} [\sigma_\nu]_{\sigma_2\sigma_1}
\Pi_{\sigma'_1\sigma'_2,\sigma_2\sigma_1}({\bf q}+{\bf q}_{\sigma_1\sigma_2}, \omega) 
\Bigr|_{{\bf q}_{\sigma_1\sigma_2} \equiv {\bf q}_{\sigma'_1\sigma'_2}}, 
\end{equation}
where $\sigma_{\mu}$ and $\sigma_{\nu}$ are the Pauli matrices, and 
${\bf q}_{\sigma_1\sigma_2} \equiv {\bf k}_{\sigma_1} - {\bf k}_{\sigma_2}$. 
Summation in spin is restricted to the cases when ${\bf q}_{\sigma'_1\sigma'_2}$ 
and ${\bf q}_{\sigma_1\sigma_2}$ are equivalent up to reciprocal-lattice translations. 
This restriction is necessary, because neutron scattering experiments observe the components 
whose absorbed and emitted momenta are equivalent~\cite{Kaneshita2001}. 
We should note that $\Pi_{\sigma'_1\sigma'_2,\sigma_2\sigma_1}({\bf q}, \omega)$ 
describes the dynamical processes where the absorbed and emitted momenta are 
${\bf q} - {\bf q}_{\sigma_1\sigma_2}$ and ${\bf q} - {\bf q}_{\sigma'_1\sigma'_2}$, 
respectively, as illustrated in Fig.~\ref{fig2}(b), where we express diagrammatically 
$\Pi_{\sigma'_1\sigma'_2,\sigma_2\sigma_1}({\bf q}, t'-t)$ as well as 
$\chi_{\sigma'_1\sigma'_2,\sigma_2\sigma_1}^{R,A}({\bf q}, \omega)$ by a solid wavy line. 

RIXS intensity from two-magnon excitations can be calculated in a similar way to that in Ref.~\onlinecite{Nomura2005} 
by means of the Keldysh diagrammatic technique~\cite{Nozieres1974}. 
In the Keldysh diagrammatic representation, probability of excitations where a pair of magnons are left in the final state 
is expressed by diagrams where a pair of magnon propagators bridge the upper normally 
and lower reversely time-ordered branches. 
As in our previous works, we adopt the Born approximation, i.e., the lowest-order perturbation 
with respect to the core-hole potential $V_{sd}$. 
Typical diagrams, which we shall calculate, 
are displayed in Fig.~\ref{fig2}(c), where a $3d$ triangle loop connects three RPA propagators, 
giving the lowest-order coupling between a charge mode (c) and two magnons (m$_1$ and m$_2$). 
These diagrams give the transition probability of the physical processes of Fig.~\ref{fig1}, 
where the charge mode excited by the $1s$ hole decays into a pair of magnons in the final state.
In Fig.~\ref{fig2}(c), we express $\chi_{\sigma_1\sigma_2,\sigma_3\sigma_4}^C({\bf q}, \omega)$ 
as well as $\Pi_{\sigma_1\sigma_2,\sigma_3\sigma_4}({\bf q}, \omega)$ by a solid wavy line. 
As known from Fig.~\ref{fig2}(c), RIXS intensity from two-magnon excitations consists of two contributions: 
\begin{eqnarray}
W(q{\bf e}; q'{\bf e}') = W_p(q{\bf e}; q'{\bf e}') + W_c(q{\bf e}; q'{\bf e}'), 
\end{eqnarray}
where the left [right] diagram including a parallel [crossed] pair of magnon propagators expresses 
$W_p(q{\bf e}; q'{\bf e}')$ [$W_c(q{\bf e}; q'{\bf e}')$]. 
$q=({\bf q}, \omega)$ and ${\bf e}$ [$q'=({\bf q}', \omega')$ and ${\bf e}'$] 
are the four-momentum and polarization of the absorbed [emitted] x-ray. 
Analytic expressions of these diagrams are given in the form of the convolution of the dynamical correlation function: 
\begin{widetext}
\begin{eqnarray}
W_p(q{\bf e}; q'{\bf e}') &=& \sum_{\sigma_i,\sigma'_i} \frac{1}{N} \sum_{\bf p} \int_{-\infty}^{\infty} \frac{d \zeta}{2 \pi} 
V_{\sigma_1\sigma_2,\sigma_3\sigma_4}({\bf e}, {\bf e}'; \omega; Q; p) 
\Pi_{\sigma'_2\sigma'_1,\sigma_1\sigma_2}({\bf p}, \zeta)  \nonumber \\ 
&&\times \Pi_{\sigma'_4\sigma'_3,\sigma_3\sigma_4}({\bf Q}-{\bf p}, \Omega-\zeta) 
V_{\sigma'_1\sigma'_2,\sigma'_3\sigma'_4}^*({\bf e}, {\bf e}'; \omega; Q; p), \label{eq:wp} \\
W_c(q{\bf e}; q'{\bf e}') &=& \sum_{\sigma_i,\sigma'_i} \frac{1}{N} \sum_{\bf p} \int_{-\infty}^{\infty} \frac{d \zeta}{2 \pi} 
V_{\sigma_1\sigma_2,\sigma_3\sigma_4}({\bf e}, {\bf e}'; \omega; Q; p) 
\Pi_{\sigma'_2\sigma'_1,\sigma_1\sigma_2}({\bf p}, \zeta)  \nonumber \\ 
&&\times \Pi_{\sigma'_4\sigma'_3,\sigma_3\sigma_4}({\bf Q}-{\bf p}, \Omega-\zeta) 
V_{\sigma'_3\sigma'_4,\sigma'_1\sigma'_2}^*({\bf e}, {\bf e}'; \omega; Q; Q-p), \label{eq:wc}
\end{eqnarray}
\end{widetext}
where $Q \equiv ({\bf Q}, \Omega) \equiv q-q' \equiv ({\bf q} - {\bf q}', \omega - \omega')$ is the momentum transfer 
and energy loss of x-rays, $p=({\bf p}, \zeta)$, $V_{\sigma_1\sigma_2,\sigma_3\sigma_4}({\bf e}, {\bf e}'; \omega; Q; p)$ 
is the scattering function expressed by the product of the $1s$-$4p$ triangle loop (the resonance factor), 
the core-hole potential screened by the charge mode c, and the $3d$ triangle loop: 
\begin{equation}
V_{\sigma_1\sigma_2,\sigma_3\sigma_4}({\bf e}, {\bf e}';\omega; Q; p) = 
R({\bf e}, {\bf e}'; \omega; \Omega) V_{sd} \sum_{\tau_3\tau_4} \Lambda_{\tau_3\tau_4}({\bf Q}, \Omega) 
L_{\tau_3\tau_4; \sigma_1\sigma_2,\sigma_3\sigma_4}(Q; p), \label{eq:v}
\end{equation}
where 
\begin{eqnarray}
R({\bf e}, {\bf e}'; \omega; \Omega) &=& 2 \sum_{\mu} \frac{1}{N} \sum_{\bf k} 
\frac{w_{\mu}({\bf q},{\bf e}) w_{\mu}^*({\bf q}',{\bf e}')} 
{[\omega + \varepsilon_{1s} + i \Gamma_{1s} - \varepsilon_{4p\mu}({\bf k})] 
[\omega' + \varepsilon_{1s} + i \Gamma_{1s} - \varepsilon_{4p\mu}({\bf k})]}, \\
\Lambda_{\tau_3\tau_4}({\bf Q}, \Omega) &=& \delta_{\tau_3\tau_4} 
- \sum_{\sigma,\tau_1\tau_2} \Gamma_{\tau_1\tau_3,\tau_4\tau_2}^{(0)} 
\chi_{\tau_1\tau_2, \sigma\sigma}^C({\bf Q}, \Omega), \label{eq:lamb}\\
L_{\tau_3\tau_4; \sigma_1\sigma_2,\sigma_3\sigma_4}(Q; p) &=& \sum_{\kappa_i} 
\frac{1}{N} \sum_{\bf k} \int_{-\infty}^{\infty} \frac{d \xi}{2 \pi} 
G_{\kappa_4\tau_3}^C({\bf k+p}, \xi + \zeta) 
G_{\tau_4\kappa_1}^C({\bf k+p-Q}, \xi + \zeta - \Omega) \nonumber \\
&& \times G_{\kappa_2\kappa_3}^C({\bf k}, \xi) 
\Gamma_{\sigma_1\kappa_3,\kappa_4\sigma_2}^{(0)} 
\Gamma_{\kappa_1\sigma_3,\sigma_4\kappa_2}^{(0)}, \label{eq:loop}
\end{eqnarray}
with 
\begin{equation}
G_{\sigma_1\sigma_2}^C({\bf k}, \xi) = \sum_j u_{\sigma_1,j}({\bf k}) 
u_{\sigma_2,j}^*({\bf k}) G_j^C({\bf k}, \xi). 
\end{equation}
$V_{sd} \Lambda_{\tau\tau}({\bf Q}, \Omega)$ corresponds to the screened core-hole potential, 
which is reduced from the bare $V_{sd}$ due to electron-hole excitations 
$\chi_{\tau_1\tau_2,\sigma\sigma}^C({\bf Q}, \Omega)$. 
Energy-momentum integration in Eq.~(\ref{eq:loop}) is numerically performed using FFT. 
In terms of the density of states (DOS) of the $4p_\mu$ band, 
\begin{equation}
\rho_{4p\mu}(\varepsilon) = \frac{1}{N} \sum_{\bf k} \delta(\varepsilon - \varepsilon_{4p\mu}({\bf k})), 
\end{equation}
the resonance factor is written as 
\begin{eqnarray}
R({\bf e}, {\bf e}'; \omega; \Omega) = 2 \sum_{\mu} \int_0^{\infty} d\varepsilon 
\frac{w_{\mu}({\bf q},{\bf e}) w_{\mu}^*({\bf q}',{\bf e}') 
\rho_{4p\mu}(\varepsilon)}{[\omega + \varepsilon_{1s} + i \Gamma_{1s} - \varepsilon] 
[\omega + \varepsilon_{1s} + i \Gamma_{1s} - \varepsilon - \Omega)]}. 
\label{eq:resfac}
\end{eqnarray}
For the below numerical calculations for RIXS spectra, we use $\rho_{4p\mu}(\varepsilon)$ 
obtained from a first-principles band calculation and $\varepsilon_{1s} = -8980$ eV, 
which reproduce well experimental x-ray absorption spectra (XAS) as shown in the Appendix. 
Rigorously speaking, the expressions (\ref{eq:wp}) and (\ref{eq:wc}) include not only two-magnon excitations but also 
purely charge excitations. However, as far as we see low-energy excitations below 1 eV, we may regard the two-magnon excitations 
completely dominate the spectral weights, since charge modes below the insulating gap ($\sim 2.5$ eV) 
should be absent in the final state. 

The above expression of RIXS intensity contains not only magnon propagators, but the resonance factor and the $3d$ triangle loop. 
To see only the the contribution from two-magnon excitations, we define the bare part of two-magnon excitations: 
\begin{widetext}
\begin{equation}
\tilde{W}({\bf Q}, \Omega) = \sum_{\sigma_i,\sigma'_i} \frac{1}{N} \sum_{\bf p} \int_{-\infty}^{\infty} \frac{d \zeta}{2 \pi} 
\Pi_{\sigma'_2\sigma'_1,\sigma_1\sigma_2}({\bf p}, \zeta) \Pi_{\sigma'_1\sigma'_2,\sigma_2\sigma_1}({\bf Q}-{\bf p}, \Omega-\zeta), \label{eq:wt}
\end{equation}
\end{widetext}
which is diagrammatically represented in Fig.~\ref{fig2}(d). 
$\tilde{W}({\bf Q}; \Omega)$ is obtained from Eq.~(\ref{eq:wp}) by maintaining total-spin conservation 
but neglecting the momentum-frequency and polarization dependences 
in $V_{\sigma_1\sigma_2,\sigma_3\sigma_4}({\bf e}, {\bf e}'; \omega; Q; p)$. 

Finally, we define the integrated weights: 
\begin{eqnarray}
I({\bf e}, {\bf e}'; {\bf Q})         & \equiv & C \int_0^{E_c} W(q{\bf e};q'{\bf e}') d \Omega , \\
\tilde{I}({\bf Q}) & \equiv & \tilde{C} \int_0^{E_c} \tilde{W}({\bf Q}, \Omega) d \Omega,  
\end{eqnarray}
where we choose $E_c = 1$ eV, and the scaling constants 
$C$ and $\tilde{C}$ to satisfy $I({\bf e}, {\bf e}'; {\bf Q}) = \tilde{I}({\bf Q}) = 1$ at ${\bf Q}=(\pi,0)$, 
for the following numerical calculations. 

\section{Numerical Results}
\label{sc:Numerical Results}

\subsection{Single-magnon excitations}
\label{sbsc:Single-magnon excitations}

\begin{figure}[htb]
\includegraphics[width=0.6\columnwidth]{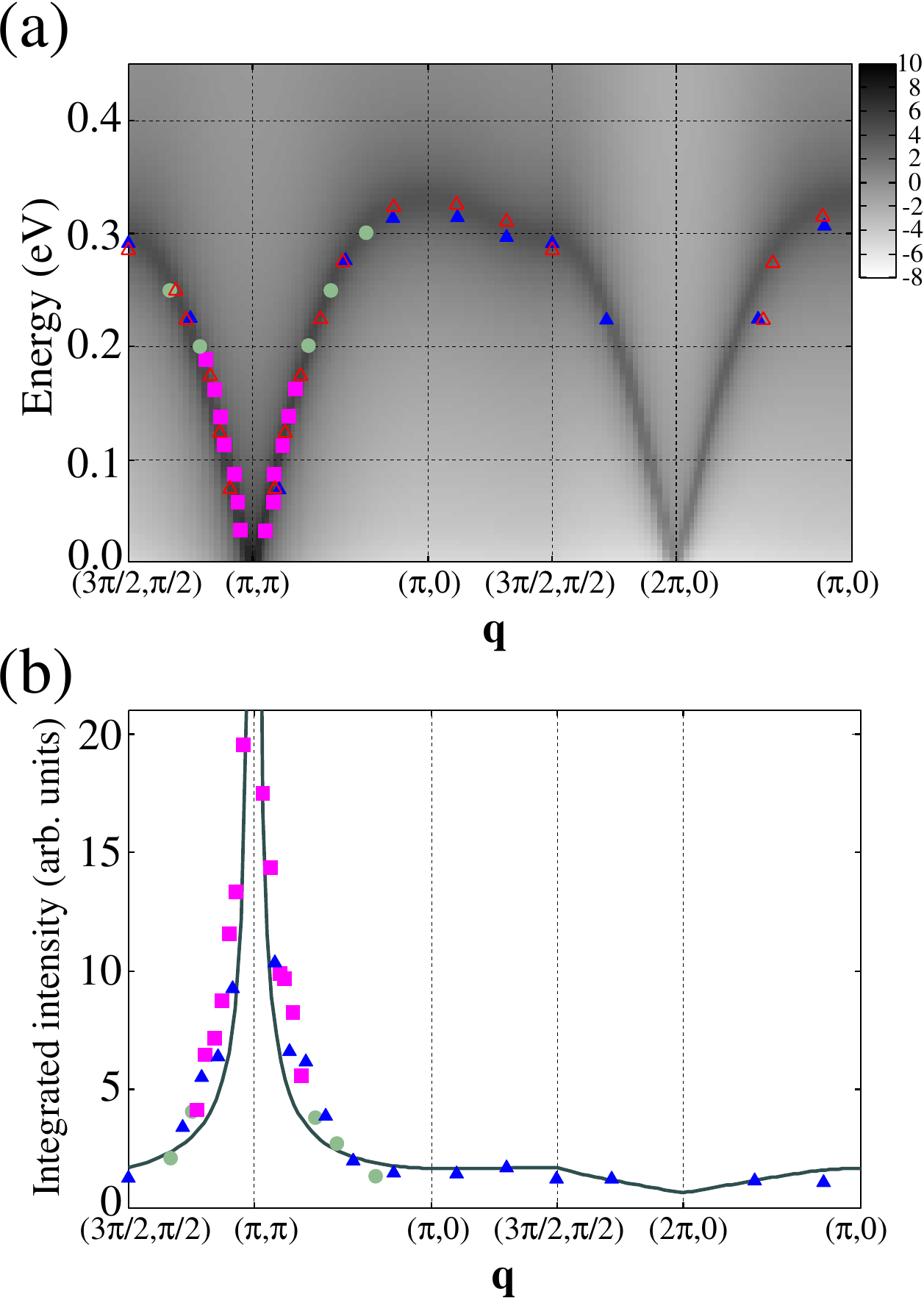}
\caption{(Color online)
(a) Calculated dynamical spin correlation function $S_{zz}({\bf q}, \omega)$ along symmetry lines 
is represented by the gray-level map in a logarithmic scale. 
(b) Integrated intensity along symmetry lines is represented by a curve, 
where $S_{zz}({\bf q}, \omega)$ is integrated in $\omega$ up to 1 eV. 
In the both panels, plots are the neutron-scattering data read from Ref.~\onlinecite{Coldea2001}: 
$T$ = 10 K (empty symbols) and 295 K (filled symbols). 
Squares are obtained for $E_i$=250 meV, circles for $E_i$ = 600 meV, 
and triangles for $E_i$ = 750 meV, where $E_i$ is incident neutron energy~\cite{Coldea2001}.}
\label{fig3}
\end{figure}

Consistency between calculated single-magnon spectra and neutron scattering data 
is prerequisite for calculation of two-magnon RIXS spectra. 
Numerical results for the dynamical spin correlation function 
$S_{zz}({\bf q}, \omega)$ are displayed in Fig.~\ref{fig3}(a). 
As shown in Ref.~\onlinecite{Peres2002}, the SDW approach yields a spin-wave 
(i.e., single-magnon) dispersion relation precisely agreeing with neutron scattering. 
Although we use the $d$-$p$ model which includes the O-$2p$ orbitals and differs 
from the simple Hubbard model in Ref.~\onlinecite{Peres2002}, 
the dispersion is well reproduced again to agree with neutron scattering data. 
It is known that SW calculation in the simple nearest-neighbor Heisenberg model 
leads to $\omega(\pi,0) = \omega(\pi/2,\pi/2)$, being inconsistent with neutron scattering. 
An advantage of the SDW formalism is that the magnon excitation energies are 
correctly reproduced at those {\bf q} points. 
One of notable features is that the strong divergent behavior around ${\bf q}=(\pi,\pi)$, 
as observed in neutron scattering~\cite{Coldea2001}, 
which is clearly seen in terms of the integrated intensity in Fig.~\ref{fig3}(b). 
This divergent behavior becomes important for the two-magnon RIXS intensity around the zone center, 
as we shall see below. 

In SW approaches, magnon DOS can be calculated straightforwardly, 
since the dispersion relation of magnons is expressed explicitly in terms of trigonometric functions. 
On the other hand, in our SDW approach, magnon DOS is difficult to calculate precisely, 
since the magnon energies are obtained only as numerical values and the calculated magnon peaks 
in $S_{\mu\nu}({\bf q},\omega)$ are accompanied by some broadness (i.e., damping). 

\subsection{Dependence on x-ray momentum transfer}

\begin{figure}[htb]
\includegraphics[width=0.7\columnwidth]{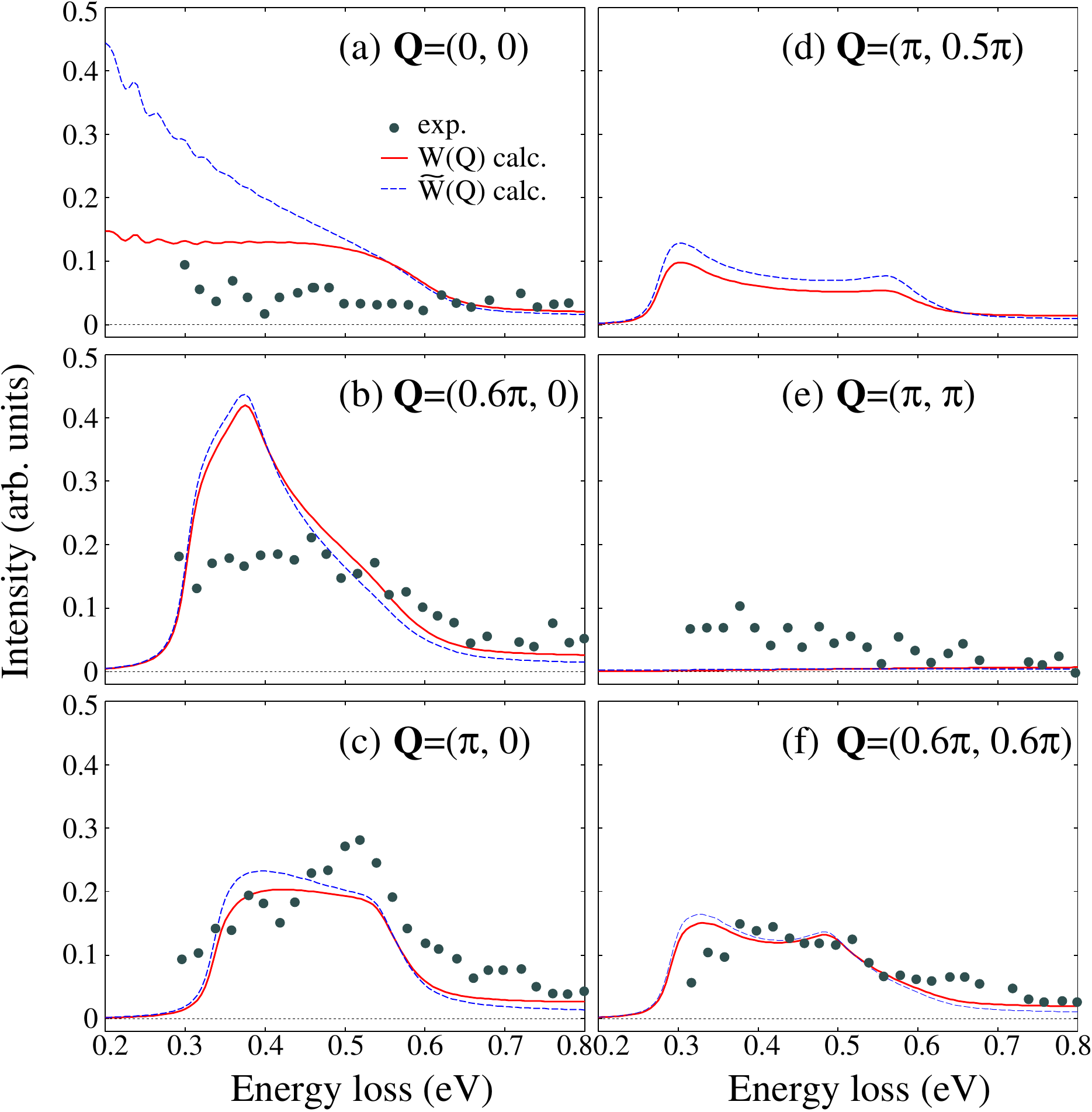}
\caption{(Color online)
Comparison with experimental data at various momentum transfers. 
Solid and broken curves are the calculated results of $W(q{\bf e}; q'{\bf e}')$ 
and $\tilde{W}({\bf Q}, \Omega)$, respectively, 
where $Q \equiv ({\bf Q}, \Omega) \equiv q-q' \equiv ({\bf q} - {\bf q}', \omega - \omega')$, 
$\Omega$ is the energy loss of x-rays. 
Plots are the experimental data read from Ref.~\onlinecite{Hill2008}. 
X-ray polarization direction is fixed to $E \parallel c$. 
The calculated intensities are scaled to match the experimental plots, 
using a scale factor common to all the panels.}
\label{fig4}
\end{figure}

Calculated two-magnon RIXS spectra $W(q{\bf e}; q'{\bf e}')$ are compared with experimental data 
at various momentum transfers in Fig.~\ref{fig4}. 
In Fig.~\ref{fig4}, also the bare two-magnon spectra $\tilde{W}({\bf Q}, \Omega)$ are drawn. 
In Fig.~\ref{fig4}(e), vanishing of the two-magnon RIXS spectrum at ${\bf Q}=(\pi,\pi)$ is reproduced, 
being consistent with experiments and also with other previous theoretical calculations. 
Considering uncertainty due to elastic-line subtraction and limited resolution in the experiment, 
we can regard the theoretical curves as agree well with the experimental plots. 

\begin{figure}[htb]
\includegraphics[width=0.8\columnwidth]{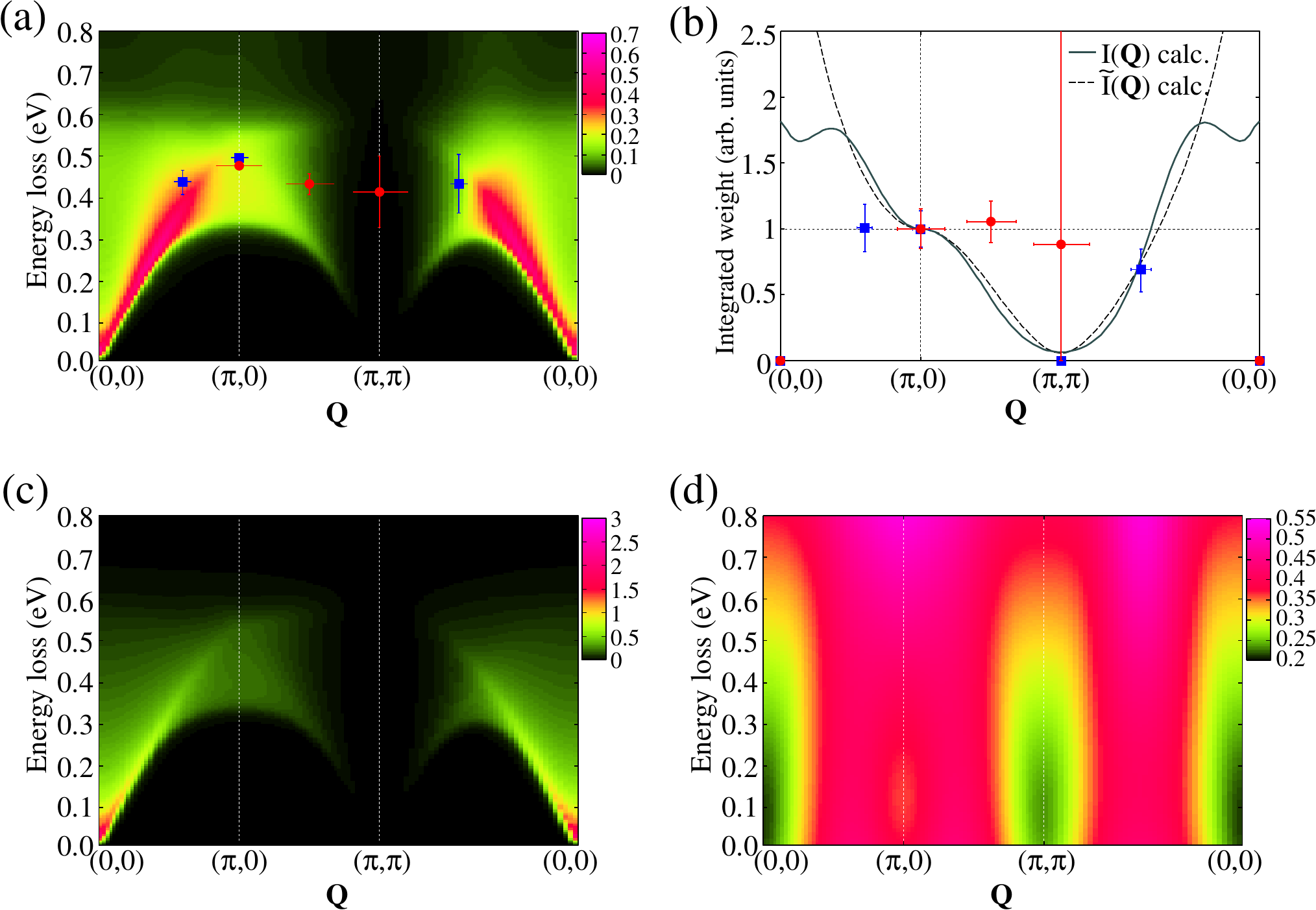}
\caption{(Color online)
(a) Calculated RIXS intensity $W(q{\bf e}; q'{\bf e}')$ and experimental peak positions along symmetry lines. 
Plots are experimental data read from Fig.~3(a) of Ref.~\onlinecite{Ellis2010}. 
(b) Integrated weights: solid and broken curves are 
the calculated results of $I({\bf e},{\bf e}'; {\bf Q})$ and $\tilde{I}({\bf Q})$, respectively, 
and plots are the experimental data read from Fig.~3(b) of Ref.~\onlinecite{Ellis2010}. 
The weights are normalized by the value at ${\bf Q}=(\pi,0)$. 
For the experimental data, background is subtracted so that the spectral weight 
at the zone center is zero~\cite{Ellis2010}. 
In (a) and (b), two symbols, circle and square, represent two different data sets at $T=300$~K and $T = 45$~K 
from two different beamlines~\cite{Ellis2010}. 
Horizontal bars of plots represent the {\bf Q} resolution~\cite{Ellis2010}. 
In the calculated results of (a) and (b), x-ray polarization direction is always fixed to $E \parallel c$, 
and the incident x-ray energy for calculation is fixed to $\omega=8992$ eV. 
(c) Bare two-magnon part, $\tilde{W}({\bf Q}, \Omega)$, calculated by Eq.~(\ref{eq:wt}). 
(d) $|\Lambda_{\tau\tau}({\bf Q}, \Omega)|$ calculated by Eq.~(\ref{eq:lamb}) along symmetry lines.} 
\label{fig5}
\end{figure}

To see more detailed momentum dependence of the two-magnon RIXS spectrum, 
we present the intensity map along symmetry lines in the first Brillouin zone 
in Fig.~\ref{fig5}(a), where we compare with experimental peak positions. 
Although the calculated intensity map shows broadness of spectra, the experimental peak positions 
fall well within an energy-loss range where the calculated intensity is relatively strong, 
except for ${\bf Q}=(\pi,\pi)$. 

A notable feature is that the RIXS intensity becomes strong around ${\bf Q}=0$ and $\Omega=0$, 
which has not been pointed out in previous experiments and theoretical calculations. 
This notable feature is straightforwardly understood in the following way: 
Intensity of single-magnon excitations becomes divergent toward ${\bf q}=(\pi,\pi)$ 
as observed in neutron scattering and also as calculated in \ref{sbsc:Single-magnon excitations}. 
Therefore, excitation of two magnons with ${\bf p}_1, {\bf p}_2 \approx (\pi,\pi)$ contributes 
strongly to the two-magnon spectrum around ${\bf Q} = {\bf p}_1 + {\bf p}_2 \approx (2\pi,2\pi) \equiv (0,0)$. 
This is a striking difference from the previous two-magnon RIXS calculations. 
The difference arises from the difference in starting models, rather than from that 
in the underlying microscopic mechanisms of two-magnon creation. 
As shown in Ref.~\onlinecite{VanDenBrink2007}, the magnetic scattering operator involving two spins 
is proportional to or commutable with the starting Heisenberg spin Hamiltonian at ${\bf Q}=0$, 
and therefore can never excite the ground eigenstate to any other states. 
This peculiarity of Heisenberg spin models leads to the vanishing of RIXS intensity at ${\bf Q}=0$. 
On the other hand, such vanishing does not occur in Hubbard-type Hamiltonians. 

In Fig.~\ref{fig5}(b), we compare the integrated weights $I({\bf e}, {\bf e}'; {\bf Q})$ 
and $\tilde{I}({\bf Q})$ with experimental data. 
Integration of the intensities yields a large weight near the zone center, 
which contradicts the vanishing weight concluded in previous theoretical and experimental works. 
In Heisenberg spin models, the intensity vanishes at the zone center and therefore also the integrated weight vanishes there, 
unless ring spin exchange is taken into account~\cite{Nagao2007,VanDenBrink2007,Forte2008}. 
As we pointed out already, this vanishing arises from the peculiarity of Heisenberg spin models. 
In experiment, the RIXS intensity around the zone center is overlaid by the much stronger elastic line, 
and is difficult to distinguish from it. 
In Ref.~\onlinecite{Ellis2010}, the two-magnon RIXS weight is set to zero at the zone center after background subtraction. 
However, supposing that two-magnon RIXS intensity may be subtracted together with the elastic line in analyzing experimental data, 
still one cannot exclude the possibility that the two-magnon RIXS weight takes a finite value at the zone center. 

In Fig.~\ref{fig5}(b), the calculated weight at ${\bf Q}=(\pi, \pi)$ seems finite, 
which will be inaccurate due to the limited precision of numerical integration. 
One may expect that the weight approaches zero more closely, 
if we take smaller $\gamma$ and finer energy-momentum discretization for numerical integration. 

In Fig.~\ref{fig5}(c), the intensity map of the bare part of two-magnon excitations, 
$\tilde{W}({\bf Q}, \Omega)$, is presented along symmetry lines in the first Brillouin zone. 
From comparison between $W(q{\bf e}; q'{\bf e}')$ and $\tilde{W}({\bf Q}, \Omega)$ 
in Fig.~\ref{fig4}, Figs.~\ref{fig5}(a) and (c), the bare two-magnon part $\tilde{W}({\bf Q}, \Omega)$ 
well captures the overall spectral properties of RIXS spectra $W(q{\bf e}; q'{\bf e}')$, 
except around the zone center ${\bf Q}=0$. 

\subsection{Dependence on incident x-ray energy and polarization}
\label{sbsc:Polarization}

\begin{figure}[htb]
\includegraphics[width=0.6\columnwidth]{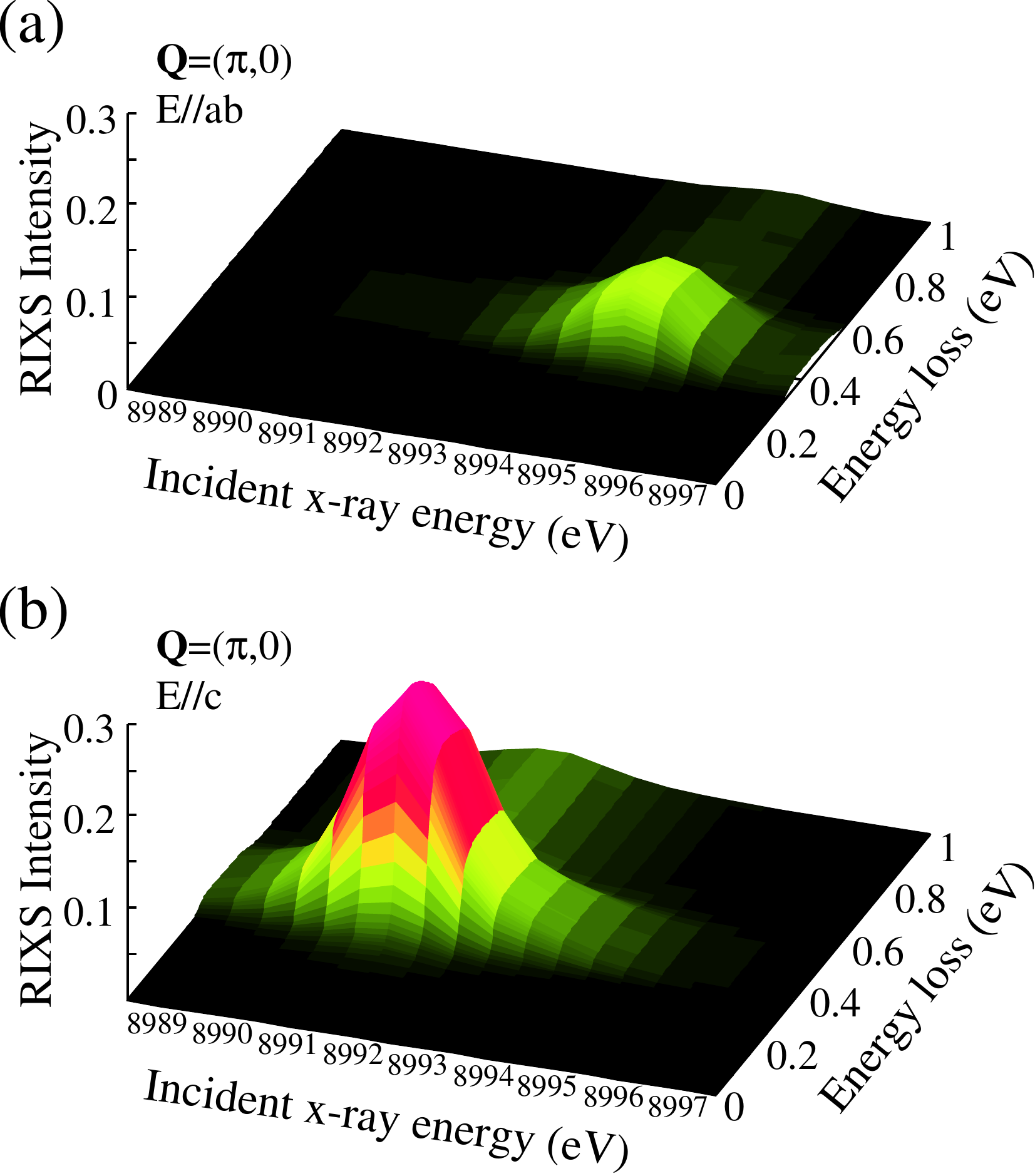}
\caption{(Color online)
Incident x-ray energy dependence for two polarization directions 
(a) $E \parallel ab$ and (b) $E \parallel c$. 
X-ray momentum transfer is fixed to ${\bf Q} = (\pi, 0)$.}
\label{fig6}
\end{figure}

In experiments~\cite{Hill2008,Ellis2010}, the two-magnon RIXS weight 
was not observed for the polarization geometry $E \parallel ab$. 
To explain this, we calculate the dependence of two-magnon RIXS spectra 
on x-ray polarization direction, whose results are shown in Fig.~\ref{fig6}. 
According to the calculated results, the two-magnon RIXS intensity for $E \parallel ab$ 
will be enhanced at higher incident x-ray energies (by about 4 eV) than that for $E \parallel c$. 
However, the intensity for $E \parallel ab$ is much weaker even at the resonance 
than for $E \parallel c$, which explains the missing of this feature 
for $E \parallel ab$ in experiments. 
This weakness of the intensity for $E \parallel ab$ arises from the smallness of the resonance factor, 
and therefore, roughly speaking, can be attributed to the smallness and broadness 
of the $4p$ DOS $\rho_{4px,y}(\varepsilon)$ (see Fig.~\ref{figA1}). 
More careful investigations may confirm a small but finite two-magnon weight for $E \parallel ab$. 

\section{Discussion and Summary}

Comparing the RIXS intensity $W(q{\bf e}; q'{\bf e}')$ and bare two-magnon part $\tilde{W}({\bf Q}, \Omega)$ is quite illuminating. 
From Fig.~\ref{fig4}, calculated $W(q{\bf e}; q'{\bf e}')$ and $\tilde{W}({\bf Q}, \Omega)$ exhibit 
quite similar spectral properties except around the zone center ${\bf Q}=0$. 
This similarity suggests that, except around ${\bf Q}=0$, 
the spectral properties are determined almost only by the two-magnon states in the final state, 
and do not depend on the details of the momentum-frequency dependences of scattering function 
$V_{\sigma_1\sigma_2,\sigma_3\sigma_4}({\bf e}, {\bf e}'; \omega; Q; p)$. 
On the other hand, difference around ${\bf Q}=0$ is striking, which is more clearly seen in the integrated weights 
$I({\bf e}, {\bf e}'; {\bf Q})$ and $\tilde{I}({\bf Q})$ in Fig.~\ref{fig5}(b): 
$I({\bf e}, {\bf e}'; {\bf Q})$ is much suppressed around ${\bf Q}=0$. 
This suppression around ${\bf Q}=0$ arises from the momentum dependence of $\Lambda_{\tau\tau'}({\bf Q},\Omega)$. 
As we mentioned,  $\Lambda_{\tau\tau'}({\bf Q},\Omega)$ reflects the strength 
of core-hole screening due to the electron-hole excitation mode with momentum {\bf Q}. 
We show $|\Lambda_{\tau\tau}({\bf Q},\Omega)|$ in Fig.~\ref{fig5}(d), 
where the smallness of $|\Lambda_{\tau\tau}({\bf Q},\Omega)|$ around ${\bf Q}=0$ indicates 
that the core-hole potential is well screened to be effectively weak there. 
Such an effect arising from local core-hole screening is characteristic of RIXS, and absent in Raman light scattering. 

As we explained already, the strong intensity around ${\bf Q}=0$ is not obtained from Heisenberg spin models. 
Furthermore, we point out that this strong intensity around ${\bf Q}=0$ cannot be captured 
by the FL framework on Raman light scattering. 
The FL Hamiltonian Eq.~(\ref{eq:FL}) leads to a momentum dependent scattering vertex 
as a result from the nature of inter-site interactions, and fails to pick up 
the strong contribution from the two magnons ${\bf p}_1, {\bf p}_2 \approx (\pi, \pi)$. 
In fact, the dominant B$_{1g}$ Raman spectrum is calculated by integrating a function 
including a factor $\sim \cos p_x - \cos p_y$~\cite{Chubukov1995,Sandvik1998}, 
which makes the above two magnons ${\bf p}_1, {\bf p}_2$ much less important. 

In Raman light scattering, characteristic spectral lineshapes confirmed the importance 
of magnon-magnon interactions~\cite{Elliott1968,Fleury1968b,Elliott1969,Parkinson1969,Fleury1970}. 
Raman light scattering spectrum can be related to the ${\bf Q}=0$ weight, 
since both the absorbed and emitted rays take a negligibly small wavenumber. 
Therefore, as in Raman light scattering, magnon-magnon interactions may affect the RIXS intensity 
at the zone center ${\bf Q} = 0$. 
However, in our calculation on RIXS, magnon-magnon interactions are not included. 
In our framework, magnon-magnon interactions are expressed by diagrams in which the two wavy lines, m$_1$ and m$_2$, 
in Fig.~\ref{fig2}(c) are connected by some diagram elements. 
Unfortunately, magnon-magnon interactions are difficult to include with some closed form and feasible numerical computation. 
On the other hand, it is still controversial whether magnon-magnon interactions are important for ${\bf Q} \neq 0$ or not. 
One reason why magnon-magnon interactions may not be effective for ${\bf Q} \neq 0$ is that two magnons with finite ${\bf Q}$ 
have relative non-zero velocity and rapidly go away from each other, and therefore the interaction 
between them is effectively diminished~\cite{Ellis2010,Vernay2007}. 

In summary, we have calculated the two-magnon RIXS spectra at the $K$-edge, 
using an itinerant Hubbard-type Hamiltonian and the SDW mean-field formalism. 
Single-magnon excitation has been described within RPA. 
Coupling between the $K$-shell hole and the magnons in the intermediate state has been dealt with 
by means of diagrammatic perturbation expansion in the Coulomb interaction. 
The calculated momentum and polarization dependences of two-magnon RIXS spectra agree well with those of experiments. 
A sharp contrast to previous studies based on Heisenberg spin models and the SW approximation 
is that the two-magnon excitations can take a large weight near the zone center. 
Further high-resolution experiments around the zone center may provide us with insights on: 
which of the itinerant Hubbard-type models and the Heisenberg spin models are more appropriate 
for the description of two-magnon RIXS near the zone center, 
how strongly the core-hole potential is screened, as well as how effective the magnon-magnon interactions are. 

\acknowledgements
The author would like to thank Prof. J. Igarashi, Dr. K. Ishii and Prof. T. Nagao 
for invaluable communications.

\appendix
\section{X-ray absorption spectra}

\begin{figure}[htb]
\includegraphics[width=0.5\columnwidth]{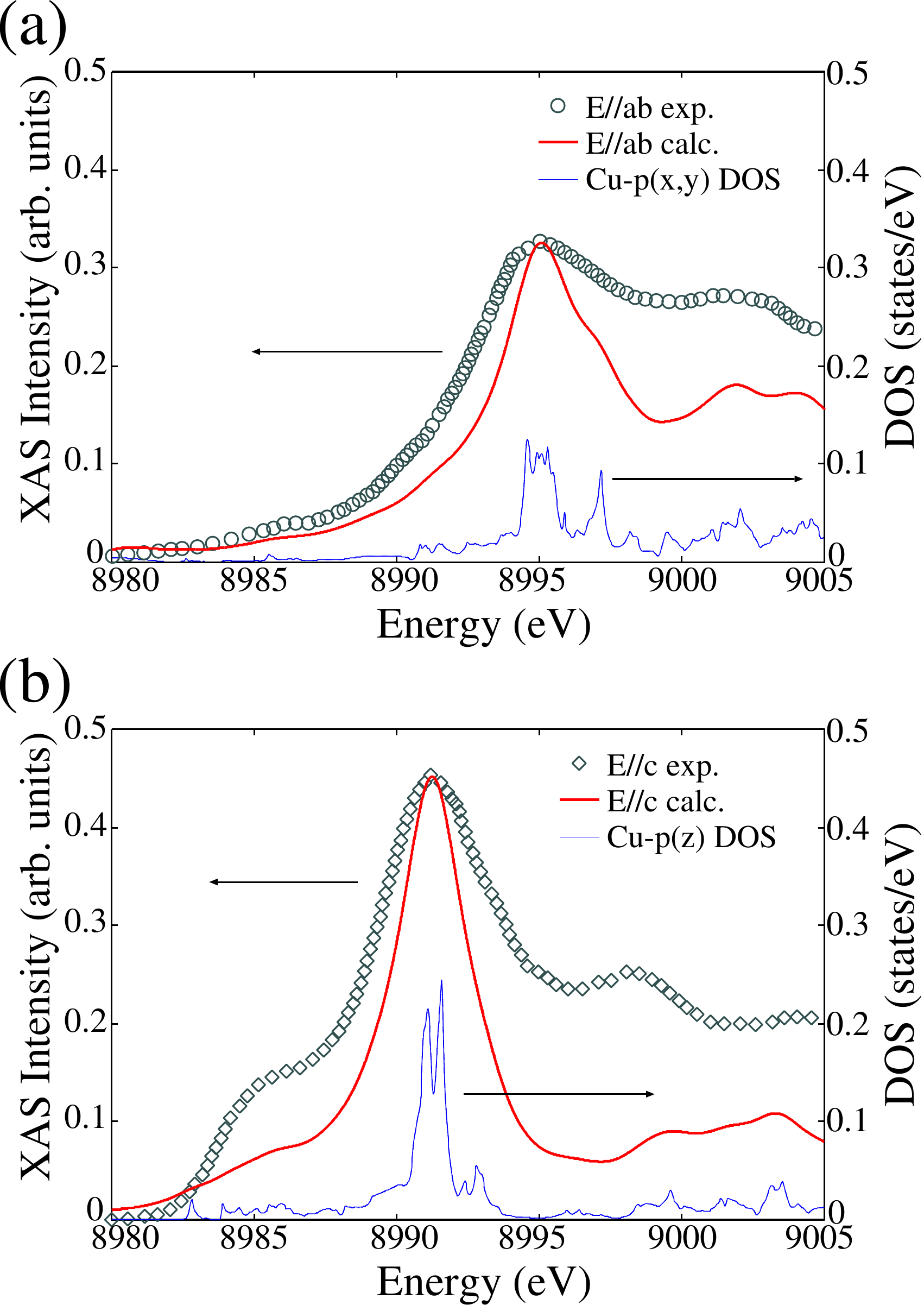}
\caption{(Color online)
Resonant x-ray absorption spectra for $E \parallel ab$ and $E \parallel c$ geometries. 
Solid and broken curves are the calculated XAS and the Cu-$p_\mu$ partial DOS 
$\rho_{4p\mu}(\omega + \varepsilon_{1s})$, respectively. 
Plots are the experimental data read from Ref.~\onlinecite{Shukla2006}. 
Experimental intensity is rescaled so that the peak intensity matches the calculated one.}
\label{figA1}
\end{figure}

X-ray absorption spectra (XAS) at the $1s$-$4p$ resonance can be approximately calculated 
from the conduction $4p$-band DOS $\rho_{4p\mu}(\varepsilon)$ with neglecting core-hole bound states: 
\begin{equation}
A(q, {\bf e}) = - 2 \sum_{\mu} |w_{\mu}({\bf q},{\bf e})|^2 \int_0^{\infty} \frac{d \varepsilon}{\pi} 
{\rm Im} \biggl[ \frac{\rho_{4p\mu}(\varepsilon) }{\omega + \varepsilon_{1s} + i \Gamma_{1s}- \varepsilon} \biggr], 
\end{equation}
where $q=({\bf q}, \omega)$, and $\Gamma_{1s} = 1$ eV in our numerical calculation. 
We calculate $\rho_{4p\mu}(\varepsilon)$ using the WIEN2k code~\cite{Blaha2015}. 
In Fig.~\ref{figA1}, calculated XAS are compared with the experimental data read from Ref.~\onlinecite{Shukla2006} 
for two polarization directions $E \parallel ab$ and $E \parallel c$. 
For the both cases of polarization, the main peak positions agree well with the experimental ones 
by setting $\varepsilon_{1s}=-8980$ eV. 


\begin{thebibliography}{99}

\bibitem{Ament2011}
L.J.P. Ament, M. van Veenendaal, T.P. Devereaux, 
J.P. Hill, and J. van den Brink, Rev. Mod. Phys. {\bf 83}, 705 (2011). 

\bibitem{Ishii2013}
K. Ishii, T. Tohyama, and J. Mizuki, J. Phys. Soc. Jpn. {\bf 82}, 021015 (2013). 

\bibitem{Hasan2000}
M.Z. Hasan, E.D. Isaacs, Z.X. Shen, L.L. Miller, 
K. Tsutsui, T. Tohyama and S. Maekawa,  
Science {\bf 288}, 1811 (2000). 

\bibitem{Kim2002}
Y.J. Kim, J. P. Hill, C.A. Burns, S. Wakimoto, R.J. Birgeneau, 
D. Casa, T. Gog and C.T. Venkataraman,  
Phys. Rev. Lett. {\bf 89}, 177003 (2002). 

\bibitem{Suga2005}
S. Suga, S. Imada, A. Higashiya, A. Shigemoto, S. Kasai, M. Sing, H. Fujiwara, 
A. Sekiyama, A. Yamasaki, C. Kim, T. Nomura, J. Igarashi, M. Yabashi and T. Ishikawa, 
Phys. Rev. B {\bf 72}, 081101(R) (2005). 

\bibitem{Ishii2011}
K. Ishii, S. Ishihara, Y. Murakami, K. Ikeuchi, K. Kuzushita, T. Inami, K. Ohwada, M. Yoshida,  
I. Jarrige, N. Tatami, S. Niioka, D. Bizen, Y. Ando, J. Mizuki, S. Maekawa and Y. Endoh, 
Phys. Rev. B {\bf 83}, 241101(R) (2011). 

\bibitem{Jarrige2012}
I. Jarrige, T. Nomura, K. Ishii, H. Gretarsson, Y.J. Kim, J. Kim, M. Upton, D. Casa, T. Gog, 
M. Ishikado, T. Fukuda, M. Yoshida, J.P. Hill, X. Liu, N. Hiraoka, K.D. Tsuei and S. Shamoto, 
Phys. Rev. B {\bf 86}, 115104 (2012). 

\bibitem{Braicovich2010}
L. Braicovich, J. van den Brink, V. Bisogni, M.M. Sala, L.J.P. Ament, N.B. Brookes, 
G.M. De Luca, M. Salluzzo, T. Schmitt, V.N. Strocov, and G. Ghiringhelli, 
Phys. Rev. Lett. {\bf 104}, 077002 (2010). 

\bibitem{Guarise2010}
M. Guarise, B. Dalla Piazza, M. Moretti Sala, G. Ghiringhelli, L. Braicovich, 
H. Berger, J. N. Hancock, D. van der Marel, T. Schmitt, V. N. Strocov, 
L. J. P. Ament, J. van den Brink, P.-H. Lin, P. Xu, H. M. R{\o}nnow, and M. Grioni, 
Phys. Rev. Lett. {\bf 105}, 157006 (2010). 

\bibitem{Tsutsui1999}
K. Tsutsui, T. Tohyama and S. Maekawa, 
Phys. Rev. Lett. {\bf 83}, 3705 (1999). 

\bibitem{Ide2000}
T. Ide and A. Kotani, 
J. Phys. Soc. Jpn. {\bf 69}, 3107 (2000). 

\bibitem{Jia2016}
C. Jia, K. Wohlfeld, Y. Wang, B. Moritz and T. P. Devereaux, 
Phys. Rev. X {\bf 6}, 021020 (2016). 

\bibitem{Nomura2004}
T. Nomura and J.I. Igarashi, 
J. Phys. Soc. Jpn. {\bf 73}, 1677 (2004). 

\bibitem{Nomura2005}
T. Nomura and J.I. Igarashi, 
Phys. Rev. B {\bf 71}, 035110 (2005). 

\bibitem{Igarashi2006}
J.I. Igarashi, T. Nomura and M. Takahashi, 
Phys. Rev. B {\bf 74}, 245122 (2006). 

\bibitem{VanDenBrink2006}
J. van den Brink and M. van Veenendaal, 
Europhys. Lett. {\bf 73}, 121 (2006). 

\bibitem{Ament2007}
L.J.P. Ament, F. Forte, and J. van den Brink, 
Phys. Rev. B {\bf 75}, 115118 (2007). 

\bibitem{Pakhira2012}
N. Pakhira, J.K. Freericks and A.M. Shvaika, 
Phys. Rev. B {\bf 86}, 125103 (2012). 

\bibitem{Hill2008}
J.P. Hill, G. Blumberg, Y.-J. Kim, D.S. Ellis, S. Wakimoto, R.J. Birgeneau, S. Komiya, Y. Ando, 
B. Liang, R.L. Greene, D. Casa, and T. Gog, Phys. Rev. Lett. {\bf 100}, 097001 (2008). 

\bibitem{Ellis2010}
D.S. Ellis, J. Kim, J.P. Hill, S. Wakimoto, R.J. Birgeneau, Y. Shvyd'ko, D. Casa, T. Gog, 
K. Ishii, K. Ikeuchi, A. Paramekanti, and Y.-J. Kim, Phys. Rev. B {\bf 81}, 085124 (2010). 

\bibitem{Nagao2007}
T. Nagao and J.I. Igarashi, Phys. Rev. B {\bf 75}, 214414 (2007). 

\bibitem{VanDenBrink2007}
J. van den Brink, Europhys. Lett. {\bf 80}, 47003 (2007). 

\bibitem{Vernay2007}
F.H. Vernay, M.J.P. Gingras, and T.P. Devereaux, Phys. Rev. B, {\bf 75}, 020403 (2007).

\bibitem{Donkov2007}
A. Donkov and A.V. Chubukov, Phys. Rev. B, {\bf 75}, 024417 (2007).

\bibitem{Forte2008}
F. Forte, L.J.P. Ament, and J. van den Brink, Phys. Rev. B {\bf 77}, 134428 (2008). 

\bibitem{Fleury1968a}
P.A. Fleury and R. Loudon, Phys. Rev. {\bf 166}, 514 (1968).

\bibitem{Elliott1968}
R.J. Elliott, M.F. Thorpe, G.F. Imbusch, R. Loudon, and J.B. Parkinson, Phys. Rev. Lett. {\bf 21}, 147 (1968).

\bibitem{Fleury1968b}
P.A. Fleury, Phys. Rev. Lett. {\bf 21}, 151 (1968).

\bibitem{Elliott1969}
R.J. Elliott and M.F. Thorpe, J. Phys. C {\bf 2}, 1630 (1969).

\bibitem{Parkinson1969}
J.B. Parkinson, J. Phys. C {\bf 2}, 2012 (1969).

\bibitem{Fleury1970}
P.A. Fleury and H.J. Guggenheim, Phys. Rev. Lett. {\bf 24}, 1346 (1970).

\bibitem{Singh1989}
R.R.P. Singh, P.A. Fleury, K.B. Lyons, and P.E. Sulewski, Phys. Rev. Lett. {\bf 62}, 2736 (1989).

\bibitem{Shastry1990}
B.S. Shastry and B.I. Shraiman, Phys. Rev. Lett. {\bf 65}, 1068 (1990).

\bibitem{Canali1992}
C.M. Canali and S.M. Girvin, Phys. Rev. B {\bf 45}, 7127 (1992). 

\bibitem{Chubukov1995}
A.V. Chubukov and D.M. Frenkel, Phys. Rev. B {\bf 52}, 9760 (1995). 

\bibitem{Schonfeld1997}
F. Sch\"onfeld, A.P. Kampf and E. M\"uller-Hartmann, Zeitschrift Phys. B {\bf 102}, 25 (1997).

\bibitem{Sandvik1998}
A.W. Sandvik, S. Capponi, D. Poilblanc and E. Dagotto, 
Phys. Rev. B {\bf 57}, 8478 (1998). 

\bibitem{Hybertsen1989}
M. S. Hybertsen and M. Schl\"{u}ter, and N.E. Christensen, Phys. Rev. B {\bf 39}, 9028 (1989).

\bibitem{Nozieres1974}
P. Nozi\`{e}res and E. Abrahams, Phys. Rev. B {\bf 10}, 3099 (1974). 

\bibitem{Takahashi2007}
M. Takahashi, J.I. Igarashi and T. Nomura, 
Phys. Rev. B {\bf 75}, 235113 (2007). 

\bibitem{Semba2008}
T. Semba, M. Takahashi and J.I. Igarashi, 
Phys. Rev. B {\bf 78}, 155111 (2008). 

\bibitem{Nomura2014}
T. Nomura, 
J. Phys. Soc. Jpn. {\bf 83}, 064707 (2014). 

\bibitem{Schrieffer1989}
J.R. Schrieffer, X.G. Wen and S.C. Zhang, Phys. Rev. B {\bf 39}, 11663 (1989).

\bibitem{Peres2002}
N.M.R. Peres and M.A.N. Ara\'{u}jo, Phys. Rev. B, {\bf 65}, 132404 (2002).

\bibitem{Zhang1988}
F.C. Zhang and T.M. Rice, Phys. Rev. B, {\bf 37}, 3759 (1988). 

\bibitem{Vaknin1987}
D. Vaknin, S.K. Sinha, D.E. Moncton, D.C. Johnston, J.M. Newsam, 
C.R. Safinya, and H.E. King, Jr., 
Phys. Rev. Lett. {\bf 58}, 2802 (1987). 

\bibitem{Kaneshita2001}
E. Kaneshita, M. Ichioka and K. Machida, J. Phys. Soc. Jpn. {\bf 70}, 866.  

\bibitem{Coldea2001}
R. Coldea, S.M. Hayden, G. Aeppli, T.G. Perring, C.D. Frost, T.E. Mason, S.-W. Cheong, and Z. Fisk, 
Phys. Rev. Lett. {\bf 86}, 5377 (2001). 

\bibitem{Blaha2015}
P. Blaha, K. Schwarz, G. Madsen, D. Kvasnicka, and J. Luitz, 
WIEN2k, An Augmented Plane Wave Plus Local Orbitals Program 
for Calculating Crystal Properties (ISBN 3-9501031-1-2). 

\bibitem{Shukla2006}
A. Shukla, M. Calandra, M. Taguchi, A. Kotani, G. Vank\'{o}, and S.-W. Cheong, 
Phys. Rev. Lett. {\bf 96}, 077006 (2006). 

\end{thebibliography}
\end{document}